\documentclass[amsfonts,12pt]{article}
\usepackage{graphicx}
\usepackage{amssymb}
\usepackage{color}

\title{Topological Bounds of Bending Energy\\ for Lipid Vesicles}
\author{Yisong Yang\\Courant Institute of Mathematical Sciences\\New York University\\New York, NY 10012, USA}
\date{}

\newcommand{\bfR}{{\Bbb R}}\newcommand{\ep}{\epsilon}

\newcommand{\sig}{{\sigma}}
\newcommand{\lm}{\lambda}\newcommand{\g}{{\mbox{g}}}
\newtheorem{oldtheorem}{Theorem}[section]
\newtheorem{oldassertion}[oldtheorem]{Assertion}
\newtheorem{oldproposition}[oldtheorem]{Proposition}

\newtheorem{oldlemma}[oldtheorem]{Lemma}
\newtheorem{olddefinition}[oldtheorem]{Definition}
\newtheorem{oldclaim}[oldtheorem]{Claim}
\newtheorem{oldcorollary}[oldtheorem]{Corollary}

\newbox\qedbox

\newcommand{\dd}{\mbox{d}}\newcommand{\sS}{{\Sigma}}
\newcommand{\ee}{\end{equation}}
\newcommand{\be}{\begin{equation}}\newcommand{\bea}{\begin{eqnarray}}
\newcommand{\eea}{\end{eqnarray}}

\newcommand{\pa}{\partial}\newcommand{\om}{\omega}

\newcommand{\nn}{\nonumber}

\begin{document}

\maketitle

\begin{abstract}
The Helfrich bending energy plays an important role in providing a mechanism for the conformation of a lipid vesicle in theoretical biophysics, which is
governed by the principle of energy minimization over configurations of appropriate topological characteristics. We will show that the presence of
a quantity called the spontaneous curvature obstructs the existence of a minimizer of the Helfrich energy over the set of embedded ring tori. Besides, despite the well-realized knowledge that lipid vesicles may present themselves in a variety of shapes of complicated topology,
there is
a lack of topological bounds for the Helfrich energy. To overcome these difficulties, we consider a general scale-invariant anisotropic curvature energy that extends the Canham elastic bending energy
developed in modeling a 
biconcave-shaped red blood cell. We will show that, up to a rescaling of the generating radii, there is a unique minimizer of the energy over the set of embedded ring tori, in the entire parameter regime, which
recovers the Willmore minimizer in its Canham isotropic limit. We also show how elevated anisotropy favors energetically a clear transition from 
spherical-, to ellipsoidal-, and then to biconcave-shaped surfaces, for a lipid vesicle. We then establish some
genus-dependent topological lower and upper bounds for the anisotropic energy. Finally, we derive  the shape equation of the generalized bending energy, which
extends the well-known Helfrich shape equation.

\medskip

{\bf Key words.} Bending energy, lipid vesicles, topological bounds, shape equations, curvatures, energy minimization
\medskip

{PACS numbers:} 02.40.Hw, 87.14.Cc, 87.16.D-, 87.16.Gj

{MSC numbers:} 53A05, 53Z05, 92C05, 92C37

\end{abstract}
 
\section{Introduction}
\setcounter{equation}{0}
\setcounter{figure}{0}

Lipid vesicles are basic life compartments confined within walls in the form of lipid membranes whose main building blocks are phospholipid molecules. A phospholipid
molecule consists of a phosphate hydrophilic head and two fatty acid hydrophobic tails. The tails of a phospholipid molecule then join those of another to
form a two-headed elementary structure known as a phospholipid bilayer. With such an elementary structure, together with a patterned assortment of other molecules such as cholesterol, proteins, and carbohydrates, a fluid mosaic \cite{SN} 
membrane in the form of a closed surface in the Euclidean space realizing a lipid vesicle, may be made to take shape which guards the interior of a cell against its exterior environment and defines the
metabolism and other life functions of the cell. 
Thus it is of  importance and interest to understand some universal properties, with regard to geometry, topology, and other mathematical and physical characteristics, of the conformation of a lipid vesicle \cite{L,G,St}.

The modeling of lipid vesicles as a subject of biophysics has a long and rich history \cite{SL,Se2}, starting from the work of Canham \cite{C} and Helfrich \cite{H}, who
use the idea of curvature energies pioneered in the earlier study of Poisson \cite{P} on surface elasticity, to give
an explanation of the biconcave shapes of red blood cells. See \cite{ZK} for a review on the understanding of cellular structures, \cite{L} for modeling
membrane conformation in biochemistry, polymer chemistry, and crystal physics, and \cite{S} for applications in solid-state nanoscale materials,
all using curvature energies, which share the common feature that the leading terms in various free energies are proportional to the total integral of the square of the mean curvature of the vesicle surface, known as the Willmore energy \cite{W,MN}. However, despite the elegant structure of the Willmore energy, it is shown \cite{Si}
that its minimum
values stay in a specific interval independent of the topology, or genus, of the surface. On the other hand, it has long been recognized \cite{L,SL,ML,Se,F,JL,MB,DE} that vesicles of lipid
bilayers may present themselves in a rich variety of geometric and topological shapes to realize a broad spectrum of life functions.
Specifically, in \cite{SL,Se,F,MB}, vesicles of toroidal as well as high-genus topology are observed.  Thus, it is imperative to identify an appropriate curvature energy, which is
consistent with the well-established curvature energies \cite{L,SL,ZK,S}, and at the same time enables us to extract information regarding geometry and topology of a
lipid vesicle. 
This will be the main motivation and  task of our present theoretical work.

The energy density that we will work with extends that of Canham \cite{C}, which is the squared sum of the two principal curvatures of the cellular surface.
In view of the Gauss--Bonnet theorem, the Canham energy differs from the Willmore energy by an integral of the Gauss curvature which is a topological invariant so that
the Canham energy is legitimately regarded as to be contained in the Helfrich energy \cite{H}. Interestingly, a close examination indicates that
the joint contributions from the Canham energy and the integral of the Gauss curvature, in the Helfrich energy, cancel out genus-dependent quantities 
and thus conceal the dependence of the total (Helfrich) energy on topology. In other words, we may take that it is the Canham energy that contains the genus-dependent
information of a lipid vesicle. Specifically, in our formalism, we shall associate the two principal curvatures with two bending rigidities in order to accommodate possible surface
anisotropy in a general setting, aimed at inclusion of a broader phenomenology. (It should be noted that the anisotropy introduced 
in our bending energy  this way is mainly motivated 
from, and limited to, a mathematical consideration of an extension of the isotropic curvature bending energy of Canham \cite{C}, as stated in Section 2. Physically realistic bending energies and actions 
modeling tethered \cite{RT1,BFT,RT2}, crystalline \cite{BT0}, nematic \cite{LMRX,XM}, and smectic \cite{SLu} membranes based on 
  anisotropic surface tension and strain tensor considerations \cite{RT1,RT2,Chen,R} could be more sophisticated  and are beyond the scope of this study.)
In fact, we will illustrate how anisotropy energetically favors ellipsoidal over
spherical and biconcave over ellipsoidal shapes
of a lipid vesicle in various parameter regimes. We then establish the anticipated genus-dependent lower and upper bounds for our free bending energy.
We will also show that the presence of a quantity called the spontaneous curvature in the Helfrich energy obstructs the existence of an energy minimizer over
the set of embedded ring tori. As an important by-product, we demonstrate that such a setback disappears with the anisotropic bending energy considered here.
In fact, we show up to a rescaling of the generating radii the existence of a unique energy minimizer among all ring tori in the full parameter regime so that it coincides with the Willmore minimizer in the
isotropic limit.
 Finally, we
derive the equation that governs the shape of the surfaces that extremize our bending energy, which extends the well-known Helfrich shape equation.

An outline of the rest of the paper is as follows. In Section 2, we introduce our scale-invariant anisotropic curvature bending energy and compare it with those of the classical Canham and
Helfrich theories. We then show how enhanced levels of anisotropy and minimum energy principle work together to drive a transition process that favors in turn
ellipsoidal and biconcave shaped surfaces over  spheres in the zero-genus situation. We also present a reinterpretation of our anisotropic energy
in view of the Helfrich energy where the spontaneous curvature is taken to be location-dependent and assume a specific form. In Section 3, we establish
some genus-dependent lower and upper bounds for our anisotropic bending energy. In particular, we show how to find and construct,
up to a rescaling, the unique energy minimizer
for our curvature bending energy over the set of embedded ring tori anywhere in the parameter regime which recovers the classical Willmore minimizer in its isotropic
limit. In Section 4, we study the minimization of the Helfrich energy and show that the direct minimization problem has no solution among the set of embedded ring tori 
except in the Willmore energy situation with a vanishing spontaneous curvature. We then show that subject to a fixed volume-to-surface-area ratio constraint,
the minimization problem of the Helfrich energy over the set of ring tori has a unique solution except for the single situation when the spontaneous curvature coincides with the constant principal curvature of the tori. Moreover, we study the Helfrich energy containing surface-area and volume contributions of the vesicle, and we show that it can be minimized over the sets of spheres and ring tori when and only when its spontaneous curvature is positive, and the radius
and radii of the energy-minimizing sphere and torus, respectively, are uniquely and explicitly determined by the coupling parameters of the bending energy. Interestingly, we
will see that the ratio of the radii of the energy-minimizing torus is bounded above universally by the critical ratio, $\frac1{\sqrt{2}}$, in the Willmore problem,
in the full parameter regime. Another surprising result is that, when we minimize the Helfrich energy containing surface-area and volume contributions over ring
tori under a fixed volume-to-surface-area ratio constraint, the spontaneous curvature obstruction to the existence of an energy minimizer disappears completely, resulting in the
acquisition of a unique least-energy torus whose ratio of generating radii may assume an arbitrary value without any restriction, so that the classical Willmore ratio $\frac1{\sqrt{2}}$ occurs in a few isolated but explicitly determined cases. As before, there is again a breakdown of the scale
invariance with the solution.
 In Section 5, we derive the shape equation of our anisotropic bending energy. This equation contains the classical
Helfrich shape equation and the Willmore equation as its special cases. In Section 6, we draw conclusions and make some comments.

\section{Free bending energies and shapes of lipid vesicles}
\setcounter{equation}{0}
\setcounter{figure}{0}

Let $k_1,k_2$ be the principal curvatures of a closed 2-surface $\Sigma$ immersed in the space $\bfR^3$ with area element $\dd \sig$.
 Recall that the two well-known competing 
bending energies modeling the deformation of a lipid membrane, in the leading orders, are the Canham energy \cite{C}
\be\label{1}
U_{\mbox{Canham}}(\Sigma)=\int_\Sigma \frac12\kappa \left(k_1^2+k_2^2\right)\,\dd \sig,
\ee
where $\kappa$ is the bending modulus, and the Helfrich energy \cite{H}
\be\label{2}
U_{\mbox{Helfrich}}(\Sigma)=\int_\Sigma \frac12\kappa \left(k_1+k_2-c_0\right)^2\,\dd \sig,
\ee
where $c_0$ is the spontaneous curvature which indicates the bending tendency of the surface. For convenience, we use dimensionless units. When $c_0=0$, (\ref{2}) is the well-known Willmore
energy in differential geometry, which has the classical lower bound $8\pi\kappa$, independent of the genus of $\sS$. In fact,
Simon \cite{Si} showed that this energy lies in the interval $[8\pi\kappa, 16\pi\kappa)$, regardless of the genus of $\sS$. One of the results of this work is to show that
the Canham energy (\ref{1}), on the other hand, enjoys a genus-dependent topological lower bound. In fact, in this work, we are interested in an extended form of
(\ref{1}), which reads
\be\label{3}
U(\Sigma)=\int_\Sigma\frac12\left(\kappa_1 k_1^2+\kappa_2 k_2^2\right)\,\dd \sig,
\ee
where the two elastic moduli, or bending rigidities, $\kappa_1,\kappa_2>0$, are present to account for possible anisotropy \cite{BT} of the lipid bilayer surface, which coincides with (\ref{1}) when isotropy is assumed so that $\kappa_1=\kappa_2$. 
Note also that the energy (\ref{3}) may remind us of the bending energy of fluid membranes studied in \cite{D,M,GK} of the form
\be\label{3b}
{\cal H}(\sS)=\int_{\sS}\left(\frac12\kappa_+\left(k_1+k_2\right)^2+\frac12\kappa_-\left(k_1-k_2\right)^2\right)\dd\sigma,
\ee
where $\kappa_+$ and $\kappa_-$ are two elastic moduli incorporating the effect of thermal fluctuations, which differs from (\ref{1})
only by a multiple of the integral of the Gauss curvature, which is invariant under surface deformation.
Furthermore
the energy (\ref{1}) or (\ref{3}) is the most natural and direct generalization of the curve bending energy in the Euler--Bernoulli elastica problem  which amounts to minimizing the
integral of the  curvature-squared of a space curve with a fixed length. Another important advantage is that  the scale invariance holds.

To proceed, using the representation of the principal curvatures $k_1,k_2$ in terms of
the mean and Gauss curvatures $H=\frac12(k_1+k_2)$ and $K=k_1 k_2$  of the surface, namely
\be
\left\{k_1,k_2\right\}=\left\{H+\sqrt{H^2-K},H-\sqrt{H^2-K}\right\},
\ee
we see that (\ref{3}) becomes
\bea\label{35}
U(\sS)&=&\om\int_\sS H^2\,\dd\sig-\frac12\om\int_\sS K\,\dd\sig\pm\delta\int_\sS H\sqrt{H^2-K}\,\dd\sig\nn\\
&\equiv&\om U^1(\Sigma)-\frac12\om U^2(\sS)\pm\delta U^3(\sS),
\eea
where $\om=\kappa_1+\kappa_2$ and $\delta=|\kappa_1-\kappa_2|$ are the sum and (absolute) difference of the elastic moduli, respectively.
In (\ref{35}), $U^1$ is of course the classical Willmore energy and $U^2$  the Gauss--Bonnet topological invariant,
\be\label{4}
U^2(\sS)=\int_\sS K\,\dd\sig=2\pi \chi(\sS)=4\pi(1-{\g}),
\ee
where $\chi(\Sigma)$ and $\g$ are the Euler characteristic and genus of $\Sigma$,
while $U^3$ is a new quantity taking account of the anisotropy of the bending energy.  Here the sign convention in (\ref{35}) follows the
rule that the plus sign is chosen when the greater
bending rigidity corresponds to the greater principal
curvature and the negative sign is chosen when the greater bending rigidity corresponds to the smaller principal curvature. It is clear that the
role of $U^3$ when $\delta\neq0$
works to spell out the presence of the non-umbilicity of the surface and break the democracy between the principal curvatures. Thus we see that
in  the context of the model (\ref{3}) 
a broader range of phenomenology may be achieved.

As a simple illustration, we consider the prolate spheroid $S_{a,c}$ ($c>a>0$), a degenerate ellipsoid, parametrized by
\be
{\bf x}(u,v)=(a\cos u\cos v,a\sin u\cos v,c\sin v),\quad 0\leq u\leq2\pi,\quad -\frac{\pi}2\leq v\leq\frac\pi2,
\ee
whose Gauss curvature, mean curvature, and area element are
\bea\label{8b}
K&=&\frac{c^2}{\left(a^2+[c^2-a^2]\cos^2 v\right)^2},\quad H=\frac{c\left(2a^2+[c^2-a^2]\cos^2v\right)}{2a\left(a^2+[c^2-a^2]\cos^2 v\right)^{\frac32}},\nn\\
\dd\sig&=&a\cos v\sqrt{a^2+[c^2-a^2]\cos^2 v}\,\dd u\dd v.
\eea
Since $H>0$, we consider the {\em minus} sign case in the bending energy (\ref{35}) which indicates that non-umbilicity is energetically favored.

\begin{figure}
\begin{center}
\includegraphics[height=2.5cm,width=6.6cm]{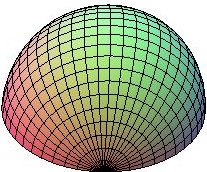}
\caption{Bending anisotropy sometimes favors an ellipsoidal over spherical shape for the vesicle (the picture shows half of the surface)}
\label{F1}
\end{center}
\end{figure}

Let $V$ denote the volume of the region enclosed by a general closed surface $\sS$, and $A$ the total surface area of $\sS$.
We are to estimate the energy (\ref{35}) for $\sS=S_{a,c}$ with a fixed volume-area ratio, $\rho=\frac VA$ (note that certain constraints such as that given by the isoperimetric inequality
may arise to prevent one from prescribing values for $V$ and $A$ freely). For this purpose, recall that 
\be
V=\frac{4\pi}3 c^3(1-\epsilon^2),\quad
A=2\pi c^2(1-\epsilon^2)\left(1+\frac1{\ep\sqrt{1-\ep^2}}\arcsin\ep\right),
\ee
 where $\ep=\sqrt{1-\frac{a^2}{c^2}}$ is the ellipticity of $S_{a,c}$.
To proceed further, we set $\rho=\frac13$, which may be realized at
\be\label{9b}
\ep=\frac{\sqrt{3}}2,\quad a=\frac c2,\quad c=\frac12\left(1+\frac{4\pi}{3\sqrt{3}}\right).
\ee
In view of (\ref{8b}) and (\ref{9b}), we have
\be\label{10b}
U^1(S_{a,c})=2\pi\left(\frac{2\sqrt{3}\pi}9+\frac54\right),\quad U^3(S_{a,c})=2\pi\left(\frac{2\sqrt{3}\pi}9-\frac14\right).
\ee
Since the volume-area ratio $\rho=\frac13$ is also enjoyed by the unit sphere $S^2$, we may use (\ref{10b}),
$U^2(S^2)=U^2(S_{a,c})$, and $U^3(S^2)=0$ to get
\be
\frac1{2\pi}\left(U(S^2)-U(S_{a,c})\right)=\om\left(2-\left[\frac{2\sqrt{3}\pi}9+\frac54\right]\right)+\delta \left(\frac{2\sqrt{3}\pi}9-\frac14\right),
\ee
which is positive when
\be\label{12b}
\frac{|\kappa_1-\kappa_2|}{\kappa_1+\kappa_2}=\frac\delta\om>\left(\frac{1-\frac{9\sqrt{3}}{8\pi}}{1-\frac{3\sqrt{3}}{8\pi}}\right)\approx 0.4787320465.
\ee
That is, subject to the volume-area ratio $\rho=\frac13$ and when (\ref{12b}) is satisfied, the prolate spheroid defined by (\ref{9b}) is energetically favored over
the unit sphere. In particular, the minimum energy surface among all the zero-genus surfaces will not be spherical. In Figure \ref{F1}, we present a plot showing the
upper half of such a favored surface.

\begin{figure}
\begin{center}
\includegraphics[height=5cm,width=10cm]{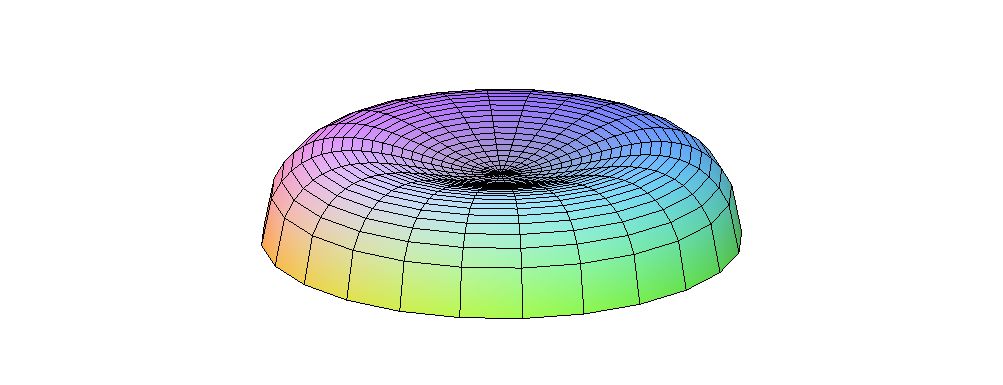}
\caption{Bending anisotropy sometimes favors a biconcave- over a spherical-shaped vesicle (the picture shows half of the surface)}
\label{F2}
\end{center}
\end{figure}

As another illustration with the same choice of the sign for the bending energy (\ref{35}), we consider an axial-symmetric surface defined by the parametrization
\be\label{14c}
{\bf x}(u,v)=(v\cos u,v\sin u, h(v)),\quad 0\leq u\leq 2\pi,\quad 0\leq v\leq R,
\ee
where $h$ is a profile function.
Then the Gauss curvature, mean curvature, and the area element of the surface (\ref{14c}) are \cite{Do,Struik}
\be\label{15b}
K=\frac{h'h''}{v\left(1+(h')^2\right)^2},\quad H=-\frac{vh''+h'(1+(h')^2)}{2v \left(1+(h')^2\right)^{\frac32}},\quad \dd\sig=v\sqrt{1+(h')^2}\,\dd u\dd v,
\ee
respectively. For our interest we study the question whether a biconcave-shaped surface may energetically be more favored over a sphere. Thus  we let our surface be defined specifically by
the biconcave-type profile function
\be
h(v)=a\left(b-(v^2-c)^2\right)^{\frac12}, \quad a,b,c>0,\quad c^2<b, \quad R=\sqrt{c+\sqrt{b}},
\ee
and we denote the surface by $\sS_0$. The fixed volume-area ratio $\rho=\frac13$ lead us to specify, e.g., $a=\frac25$ (approximate), $b=2,c=1, R=\sqrt{1+\sqrt{2}}$. Inserting these into
(\ref{15b}) and integrating by Maple $^{\mbox{\tiny TM}}$ 11, here and in the sequel, we see that the energies $U^1$ and $U^3$ have the approximate values 
\be
U^1(\sS_0)=4\pi (1.607433067),\quad U^3(\sS_0)=4\pi (0.9144890719),
\ee
which lead to
\be
\frac1{4\pi}\left(U(S^2)-U(\sS_0)\right)=-0.607433067\,\om+0.9144890719\,\delta.
\ee
Thus $U(\sS_0)<U(S^2)$ when
\be\label{18b}
\frac\delta\om>\frac{0.607433067}{0.9144890719}\approx 0.6642321769.
\ee
This condition is more stringent than (\ref{12b}).
In other words, we have seen that, when the rigidity discrepancy is sufficiently significant such that (\ref{18b}) is fulfilled, then
under the volume-area constraint, $\rho=\frac13$, a biconcave surface such as that given here and shown in Figure \ref{F2}
is energetically favored over the unit sphere.

At this moment, it will also be interesting to compare the energy of the prolate spheroid $S_{a,c}$ given in (\ref{10b}) with that of the biconcave surface $\Sigma_0$ just obtained.  In view of (\ref{10b}), we have
\bea
\frac1{4\pi} \left(U(S_{a,c})-U(\Sigma_0)\right)&=&\frac\omega{4\pi}\left( U^1(S_{a,c})-U^1(\Sigma_0)\right)-\frac\delta{4\pi}\left(U^3(S_{a,c})-U^3(\Sigma_0)\right)\nn\\
&=&-0.3778332786\,\omega +0.4348892837\,\delta,
\eea
which becomes positive when
\be\label{220}
\frac\delta\om>\frac{0.3778332786}{0.4348892837}\approx 0.8688033777.
\ee
That is, when anisotropy becomes so significant that (\ref{220}) is fulfilled, a biconcave surface will be energetically favored over a prolate spheroid under the fixed
volume-to-surface-area constraint, $\rho=\frac13$, although the latter is energetically favored over a round sphere.

We note that the afore-discussed anisotropy bounds may further be improved when the volume-to-surface-area ratio $\rho=\frac VA$ is suitably adjusted. For simplicity and clarity,
we consider the case of a prolate spheroid with $\rho=\frac1{10}$ which may be realized by setting
\be
a=\frac1{40}\left(3\sqrt{3}+2\pi\right),\quad c=\frac1{20}\left(3+\frac{2\pi}{\sqrt{3}}\right),\quad\epsilon=\frac12.
\ee
Thus we have $U^1(S_{a,c})=2\pi(2.021266457)$ and $U^3(S_{a,c})=2\pi(0.1879333913)$. On the other hand, let $S^2_r$ denote the 2-sphere of radius $r>0$. Then 
for $S^2_r$ we have $\rho=\frac r3$  so that $\rho=\frac1{10}$ gives us $r=\frac3{10}$. With this we have
\bea
\frac1{2\pi}\left(U(S^2_{r})-U(S_{a,c})\right)&=&\frac\omega{2\pi}\left(U^1(S^2_r)-U^1(S_{a,c})\right)+\frac\delta{2\pi}U^3(S_{a,c})\nn\\
&=&-0.021266457\,\omega+0.1879333913\,\delta,
\eea
which is positive when
\be
\frac\delta\om>\frac{0.021266457}{0.1879333913}\approx 0.1131595447,
\ee
which is much more relaxed than the condition (\ref{12b}). This example indicates that whenever anisotropy occurs a prolate spheroid may energetically be favored
over a round sphere provided that the ratio of volume-to-surface-area is appropriately adjusted because the bending energy (\ref{3}) is invariant with respect to the radius of
the round sphere but, on the other hand, dependent on the geometry of the prolate spheroid sensitively.

We can make a reinterpretation of the bending energy (\ref{3}) in the framework of the Helfrich energy (\ref{2}). To this end, we observe that
there holds
\be\label{13b}
\om H^2\pm\delta H\sqrt{H^2-K}=(\mu H-\nu\sqrt{H^2-K})^2 +\nu^2 K,
\ee
where
\be\label{14b}
\mu=\frac12\left(\sqrt{\om\mp\delta}+\sqrt{\om\pm\delta}\right),\quad\nu=\frac12 \left(\sqrt{\om\mp\delta}-\sqrt{\om\pm\delta}\right).
\ee
With (\ref{13b}) and (\ref{14b}), we may rewrite (\ref{35}) as
\be
U(\sS)=\frac12\kappa\int_{\sS}(2H-c_0({\bf x}))^2\,\dd\sig+\frac12\left(\frac12 \left[\sqrt{\om\mp\delta}-\sqrt{\om\pm\delta}\right]^2-\om\right)\int_{\sS}
K\,\dd\sig,
\ee
where $\kappa$ and $c_0({\bf x})$ are the effective elastic modulus and ``spontaneous curvature" given, respectively, by
\bea
\kappa&=&\frac12\mu^2=\frac18 \left(\sqrt{\om\mp\delta}+\sqrt{\om\pm\delta}\right)^2,\\
c_0({\bf x})&=&\frac{2\nu}{\mu}\sqrt{H^2-K}= 2\left(\frac{\sqrt{\om\mp\delta}-\sqrt{\om\pm\delta}}{\sqrt{\om\mp\delta}+\sqrt{\om\pm\delta}}\right)\sqrt{H^2({\bf x})-K({\bf x})}\nn\\
&=&\pm \frac{\nu}{\mu}(k_1-k_2).
\eea
It is interesting to note that, in the isotropic situation, $\delta=0$, the effective spontaneous curvature vanishes, $c_0=0$.
\section{Genus-dependent lower and upper energy bounds}
\setcounter{equation}{0}
\setcounter{figure}{0}

We now derive some topological bounds for the bending energy (\ref{3}). First, we recall the  Chern--Lashof inequality \cite{CL} 
\be\label{5}
\int_\sS|K|\,\dd\sig\geq2\pi(4-\chi(\Sigma))=4\pi(1+\g).
\ee
See also \cite{Ch,O}.

Next, using the notation $K^+=\max\{K,0\}, K^-=\max\{-K,0\}$ so that $K=K^+-K^-, |K|=K^++K^-$, $ \sS^+=\{K\geq0\}, \sS^-=\{K<0\}$, and applying (\ref{5}), we may rewrite (\ref{3}) as
\bea\label{6}
U(\sS)&=&\int_{\sS^+}\frac12\left\{(\sqrt{\kappa_1}k_1-\sqrt{\kappa_2}k_2)^2+2\sqrt{\kappa_1\kappa_2}k_1 k_2\right\}\,\dd\sig \nn\\
&&+\int_{\sS^-}\frac12\left\{(\sqrt{\kappa_1}k_1+\sqrt{\kappa_2}k_2)^2-2\sqrt{\kappa_1\kappa_2}k_1 k_2\right\}\,\dd\sig\nn\\
&=&\int_{\sS^+}\frac12(\sqrt{\kappa_1}k_1-\sqrt{\kappa_2}k_2)^2\,\dd\sig+\int_{\sS^-}\frac12(\sqrt{\kappa_1}k_1+\sqrt{\kappa_2}k_2)^2\,\dd\sig+\sqrt{\kappa_1\kappa_2}\int_\sS|K|\,\dd\sig\nn\\
&\geq&4\pi(1+\g)\sqrt{\kappa_1\kappa_2},
\eea
which is $\g$-dependent as desired. On the other hand, for (\ref{2}),  we have 
\bea\label{6b}
U_{\mbox{Helfrich}}(\sS)&=&\int_\sS\frac12\kappa (k_1^2+k_2^2)\,\dd\sig+\kappa\int_\sS K\,\dd\sig-2c_0\kappa\int_\sS H\,\dd\sig+\frac12\kappa c_0^2\int_\sS\dd\sig\nn\\
&\geq&4\pi(1+g)\kappa+4\pi(1-\g)\kappa-2c_0\kappa\int_\sS H\,\dd\sig+\frac12\kappa c_0^2\int_\sS\dd\sig\nn\\
&=&8\pi \kappa-2c_0\kappa\int_\sS H\,\dd\sig+\frac12\kappa c_0^2\int_\sS\dd\sig,
\eea
in view of (\ref{4}) and (\ref{6}), which does not render a $\g$-dependent lower bound, unfortunately.

The basic question we are interested in here is to estimate
\be\label{7}
U_{\g}=\inf\left\{ U(\sS)\,|\,\mbox{the surface $\sS$ is of genus $\g$}\right\}.
\ee
From (\ref{6}) we have the lower bound
\be\label{8}
U_\g\geq 4\pi(1+\g)\sqrt{\kappa_1\kappa_2}.
\ee

In the following, we aim to obtain some $\g$-dependent upper bounds for $U_\g$ ($\g=0,1,2,\dots$).

Case (i): $\g=0$. In this situation, we use the 2-sphere of radius $R>0$, say $S^2_R$, as a trial surface. Then $k_1=k_2=\frac1R$ so that $U(S^2_R)=2\pi (\kappa_1+
\kappa_2)$. Thus we have
\be\label{9}
\sqrt{\kappa_1\kappa_2}\leq\frac{U_0}{4\pi}\leq\frac{\kappa_1+\kappa_2}2.
\ee
That is, the quantity $\frac{U_0}{4\pi}$ lies between the geometric mean and the arithmetic mean of the bending moduli. In particular, in the isotropic limit,
$\kappa_1=\kappa_2=\kappa$ (say), $U_0=4\pi\kappa$, which is realized by all round spheres. This last statement is a classical result due to Willmore \cite{W2}.
In the anisotropic situation in which $\kappa_1\neq\kappa_2$, it is natural to expect that $U_0$  will be realized by non-round spheres such as ellipsoids or prolate or oblate spheroids,
as observed earlier. This question deserves a thorough study.

Case (ii): $\g=1$. We consider the ring torus, with the two radii, $a>b>0$, embedded in $\bfR^3$ with the standard parametrization
\be\label{3.7}
{\bf x}(u,v)=\left((a+b\cos u)\cos v,(a+b\cos u)\sin v, b\sin u\right),\quad 0\leq u, v\leq 2\pi.
\ee
We denote such a torus by $T_{a,b}^2$. The principal curvatures and area element are 
\be\label{11}
k_1=\frac1b,\quad k_2=\frac{\cos u}{a+b\cos u},\quad \dd\sig=(a+b\cos u)b\,\dd u\dd v.
\ee
Therefore
\be\label{12}
U(T_{a,b}^2)=\frac{2\pi^2}\tau\left({\kappa_1}+\kappa_2\left[\frac1{\sqrt{1-\tau^2}}-1\right]\right)\equiv 2\pi^2 f(\tau),\quad\tau=\frac ba\in (0,1).
\ee
Since $f(\tau)\to\infty$ as $\tau\to0$ and $\tau\to1$, we see that $f(\tau)$ attains its global minimum in $0<\tau<1$, for any $\kappa_1,\kappa_2>0$,
which is a root of $f'(\tau)=0$ in $(0,1)$, say $\tau_{\min}$, which in general is rather complicated, where we have
\be\label{13}
\frac{\tau^2}{\kappa_2}f'(\tau)=\frac{\tau^2}{(1-\tau^2)^{\frac32}}-\left(\gamma-1+\frac1{\sqrt{1-\tau^2}}\right)\equiv g(\tau),\quad\gamma
\equiv\frac{\kappa_1}{\kappa_2}.
\ee
From (\ref{13}), we have $g'(\tau)>0$ for $\tau\in(0,1)$. Hence $f'(\tau)=0$ has exactly one root, which is $\tau_{\min}$, in $(0,1)$, which
establishes the uniqueness of $\tau_{\min}$. By
the implicit function theorem, we see that $\tau_{\min}$ depends on $\gamma$ increasingly. Besides, it is clear that $\tau_{\min}\to 1$ when $\gamma\to\infty$ and $\tau_{\min}\to0$ when $\gamma\to 0$. As concrete examples, we have taken $\gamma$ to be
$\gamma=\frac1N$ and $\gamma=N$ ($N=1,2,\dots,100$), and we have the following sample results, which are sufficiently simple to be listed
for the pair $\left(\tau_{\min},\gamma\right)$:
\be
\left(\frac{\sqrt{19-2\sqrt{29}}}7,\frac18\right),\left(\frac1{\sqrt{2}},1\right),\left(\frac{\sqrt{\sqrt{5}-1}}{\sqrt{2}},2\right),\left(\frac{\sqrt{3}}2,5\right),\left(\frac{2\sqrt{2}}3,22\right),\left(\frac{\sqrt{15}}4,57\right),
\ee
among which $\left(\frac1{\sqrt{2}},1\right)$ is the classical result in the Willmore problem \cite{W,W2}. (In our formalism the problem asks whether $U_1\geq 4\pi^2\kappa$ when $\gamma=1$. This problem was solved by Marques and Neves \cite{MN,MN2} who also established the general bound $U_{\g}\geq 4\pi^2\kappa$
for $g\geq1$.)  Nevertheless, from solving $g(\tau)=0$ in (\ref{13}), we get
\be\label{15}
\gamma=1+\frac{2\tau_{\min}^2-1}{(1-\tau_{\min}^2)^{\frac32}},
\ee
which allows us to find $\gamma$ easily given $\tau_{\min}$. Below we list a few for the pair $\left(\tau_{\min},\gamma\right)$ again:
\be
\left(\frac12,1-\frac{4}{3\sqrt{3}}\right),
\left(\frac23,1-\frac3{5\sqrt{5}}\right),
\left(\frac34,1+\frac8{7\sqrt{7}}\right),
\left(\frac56,1+\frac{84}{11\sqrt{11}}\right).
\ee

Thus, given $\kappa_1,\kappa_2$, we can insert (\ref{15}) into (\ref{12}) to determine the minimum value of the bending energy (\ref{3}) over the set of  ring tori  to be
\be\label{17}
U_{\min}(T^2)\equiv\min\left\{U(T^2_{a,b})\,|\, a>b\right\}=2\pi^2\kappa\frac{\tau_{\min}}{(1-\tau_{\min}^2)^{\frac32}},
\ee
where $\kappa_1=\gamma\kappa,\kappa_2=\kappa$. For example, $U_{\min}(T^2)=4\pi^2 \kappa$ when $\gamma=1$, which is classical, and 
\be
U_{\min}(T^2)=2\pi^2\kappa\,\frac{4}{3\sqrt{3}}\approx 1.5396\pi^2\kappa,\quad \gamma=\left(1-\frac4{3\sqrt{3}}\right)\approx 0.2302.
\ee

In light of the Willmore problem \cite{W,W2}, it will be interesting to know whether
\be
U_1\geq U_{\min}(T^2)= 2\pi^2\kappa\frac{\tau_{\min}}{(1-\tau^2_{\min})^{\frac32}},
\ee
in anisotropic situations in which $\gamma\neq1$, and more generally, in light of \cite{MN,MN2}, whether $U_{\g}\geq U_{\min}(T^2)$ for all $\g\geq1$, again
in the situation when $\gamma
\neq1$.

Case (iii): $\g\geq2$. For simplicity, we start with $\g=2$. Let $T^2_{\min}$ be a 2-torus realizing the minimum energy $U_{\min}(T^2)$ given in (\ref{17}). Take two copies of $T^2_{\min}$,  cut an identical portion on each of them, and glue them together to obtain a smooth $\g=2$ surface, say $\tilde S$. 
We can do so in such a way that the energy of the transitional piece between the two cut-open tori which appears like a waist area may be kept smaller than
the sum of the energies of the cut-off portions of the tori. 
 Denote the undisturbed portions of and their complements  in the surfaces of the two copies of $T^2_{\min}$
by $P_1,P_2$ and $P_1',P_2'$, respectively. Use $\cal W$ to denote the waist portion of the $\g=2$ surface $\tilde S$. Then $U({\cal W})\leq U(P'_1)+U(P'_2)$.
% (cf. Figure
%\ref{F3}). 
Consequently, we have
\be
U({\tilde S})= U(P_1)+U(P_2)+U({\cal W})\leq U(P_1\cup P'_1)+U(P_2\cup P_2')=2 U_{\min}(T^2).
\ee
 Hence $U({\tilde S})\leq 2U_{\min}(T^2)$. This argument obviously allows us to establish the
general bound $U_{\g}\leq \g\, U_{\min}(T^2)$ for any $\g\geq2$.

%\begin{figure}
%\begin{center}
%\includegraphics[height=4cm,width=8cm]{double-torus}
%\caption{Two minimum-energy tori smoothly glued together to form a double torus}
%\label{F3}
%\end{center}
%\end{figure}

In summary, if we denote the unique solution $\tau_{\min}$ of the equation (\ref{15}) by $\tau(\gamma)$, then we may combine (\ref{8}), (\ref{17}), and the
above discussion to arrive at the $\g$-dependent bounds
\be\label{20}
4\pi(1+\g)\sqrt{\gamma}\,\kappa\leq U_\g\leq 2\pi^2 \g\, \kappa\,\frac{\tau(\gamma)}{(1-\tau^2(\gamma))^{\frac32}},\quad \g\geq1.
\ee

In particular, in the isotropic situation when $\gamma=1$ so that $\tau(1)=\frac1{\sqrt{2}}$, the bounds stated in (\ref{20}) assume the following elegant simple form:
\be\label{21}
4\pi(1+\g)\,\kappa\leq U_\g\leq 4\pi^2\g \,\kappa,\quad \g\geq1,
\ee
among which the case where $\g=1$ is classical.

%We note that the upper bound in (\ref{20}) may be crude and can still be improved further with some refined estimates. For clarity and simplicity, we consider the
%isotropic situation, $\kappa_1=\kappa_2=\kappa$ ($\gamma=1$). For any $\g\geq1$, we may use $\g$ half copies of the Willmore torus to grow $\g$ handles
%over a sphere, resulting in a surface $\Sigma$ with the energy bound $U(\Sigma)\leq 4\pi\kappa+2\pi^2\g\kappa$. Consequently we obtain
%\be
%U_\g\leq 2\pi(2+\g\pi)\kappa,
%\quad \g\geq1.
%\ee
%Since $4\pi^2\g>2\pi(2+\g\pi)$ if and only if 

Note that, unlike in the Willmore energy case, the energy (\ref{3}) evaluated over round spheres may exceed that over tori in some anisotropic situations. To see
this, from $U(S^2_R)=2\pi\kappa(1+\gamma)$, (\ref{12}), and (\ref{15}), we have the normalized energy difference
\bea
d(\tau)&=&\frac1{2\pi\kappa}\left(U(S^2_R)-U(T^2_{a,b})\right)\nn\\
&=&2+\frac{2\tau^2-1}{(1-\tau^2)^{\frac32}}-\frac{\pi\tau}{(1-\tau^2)^{\frac32}},\quad \tau=\tau_{\min}\in(0,1).
\eea
This quantity, as a function of $\tau$, is monotone decreasing and vanishes at $\tau_0\approx 0.2928010148$. Thus, from (\ref{15}), we see that when
$\gamma>\gamma_0$ where
\be
\gamma_0=1+\frac{2\tau_0^2-1}{(1-\tau_0^2)^{\frac32}}\approx 0.0522343632,
\ee
we have $d(\tau)<0$, namely,
the energy of a round sphere always stays below that of a torus, which includes
the well-known isotropic situation, $\gamma=1$;
on the other hand, in the extremely non-isotropic situation when $\gamma<\gamma_0$,
 $d(\tau)>0$ holds so that the energy of a round sphere stays {\em above} the minimum bending energy over the set of tori, as stated in (\ref{17}),
which seems rare and unexpected.

Note also that, although the dependence of $\tau(\gamma)=\tau_{\min}$ given in (\ref{15}) is complicated in general, nevertheless, from
\be
\frac{\dd\gamma}{\dd\tau}=\frac{\tau(1+2\tau^2)}{(1-\tau^2)^{\frac52}},\quad 0<\tau<1; \quad \lim_{\gamma\to 0}\tau(\gamma)=0,\quad
\lim_{\gamma\to\infty}\tau(\gamma)=1,
\ee
we can deduce the asymptotes
\be
\tau(\gamma)\approx \sqrt{2\gamma},\quad \gamma\to0;\quad \tau(\gamma)\approx\sqrt{1-\frac1{\gamma^{\frac23}}}\approx1-\frac1{2\gamma^{\frac23}},\quad \gamma\to\infty,
\ee
which are much simpler to use to estimate $U_\g$ in these extreme cases.

\section{Obstruction to the minimization of the Helfrich\\ energy and  its removal}
\setcounter{equation}{0}
\setcounter{figure}{0}

%Although the energy (\ref{2}) has not been shown to have a topology-dependent lower bound, we may still obtain some appropriate upper bounds. In the following, we consider the non-Willmore situation, $c_0\neq0$.

In this section, we show how the presence of the spontaneous curvature $c_0$ in (\ref{2}) obstructs the existence of an energy minimizer over the set of spheres or tori, and how such an obstruction may be removed partially or completely.

Case (i). Spheres. For $\sS=S^2_R$ ($R>0$), we see that $k_1=k_2=\frac1R$ and (\ref{2}) assumes the value
\be\label{25}
U_{\mbox{Helfrich}}(S^2_R)=2\pi\kappa\left(c_0 R-2\right)^2,
\ee
which is minimized at $R=\frac2{c_0}$ when $c_0>0$, yielding the minimum value $U_{\mbox{Helfrich}}\left(S^2_{\frac2{c_0}}\right)=0$. When $c_0<0$, 
(\ref{25}) has no minimum to attain for $R>0$, yet, it has the infimum $8\pi\kappa$. 
%Thus, if we set
%\be\label{26}
%U_{\mbox{Helfrich},\g}=\inf\left\{ U_{\mbox{Helfrich}}(\sS)\,|\,\mbox{the surface $\sS$ is of genus $\g$}\right\},
%\ee
%then we have 
%\be
%U_{\mbox{Helfrich},0}=0,\quad c_0>0; \quad U_{\mbox{Helfrich},0}\leq 8\pi \kappa,\quad c_0<0.
%\ee
When $c_0=0$, which is the classical Willmore situation, the minimum is $8\pi\kappa$, which is attained by any $S^2_R$. 

Thus, in summary for the case, we see that $c_0$ gives rise to a partial obstruction to the existence of an energy minimizer,
so that, there is non-existence when $c_0<0$, and, when $c_0>0$, although the existence is restored, there is a breakdown of the scale invariance.

Case (ii). Tori. For $\sS=T^2_{a,b}$ ($a>b>0$), applying the energy decomposition in (\ref{6b}), (\ref{11}), and (\ref{12}), we have
\be\label{28}
 U_{\mbox{Helfrich}}(T^2_{a,b})={2\pi^2\kappa}\left(\frac1{\tau\sqrt{1-\tau^2}}-2c_0 a+c_0^2 a^2\tau \right),\quad a>0,\quad\tau=\frac ba\in(0,1).
\ee
When $c_0<0$, the infimum of (\ref{28}) is attained at $\tau=\frac1{\sqrt{2}}, a=0$, which is $4\pi^2\kappa$. Thus we see that this infimum is {\em not}
attainable, that is, the Helfrich energy (\ref{2}) cannot be minimized, among the ring tori, $T^2_{a,b}$ ($a>b>0$). When $c_0>0$, we see that for fixed $\tau\in(0,1)$ the right-hand side of (\ref{28}) can be minimized at $a=\frac1{c_0\tau}$. For such a choice of $a$, we obtain from (\ref{28}) the result
\be\label{4.9}
 U_{\mbox{Helfrich}}(T^2_{a,b})={2\pi^2\kappa}\left(\frac1{\tau\sqrt{1-\tau^2}}-\frac1\tau \right)\equiv 2\pi^2\kappa\, h(\tau),\quad\tau\in(0,1),
\ee
where $h(\tau)$ is monotone increasing and $h(\tau)\to0$ as $\tau\to0$. Therefore, 
the infimum of (\ref{28}) is zero which is again not attainable among the ring tori
considered. 
%Thus we have
%\be
%U_{\mbox{Helfrich},1}\leq 4\pi^2\kappa,\quad c_0>0;\quad U_{\mbox{Helfrich},1}=0,\quad c_0<0.
%\ee

Hence, in summary for the case, we see that the infimum of the Helfrich energy (\ref{2})  is not attainable among the set of the ring tori $\{T^2_{a,b}\}_{a>b>0}$  in any non-Willmore situations in which $c_0\neq0$. This result is in sharp contrast with the Willmore situation \cite{W,MN,W2,MN2}.

%Case (iii). $\g\geq2$. For any $\vep>0$, let $T^2_{a,b}$ be such that $U_{\mbox{Helfrich}}(T^2_{a,b})<U_{\mbox{Helfrich},1}+\vep$. As earlier, we
%now glue $\g$ copies of $T^2_{a,b}$ together to form an appropriate surface, $\Sigma$, of genus $\g$, so that
%\be
%U_{\mbox{Helfrich}}(\Sigma)\leq \g U_{\mbox{Helfrich}}(T^2_{a,b})<\g\left(U_{\mbox{Helfrich},1}+\vep\right).
%\ee
%Since $\vep>0$ is arbitrary, we arrive at
%the general conclusion
%\be\label{4.11}
%U_{\mbox{Helfrich},\g}\leq 4\pi^2\g\kappa,\quad c_0>0,\quad
%U_{\mbox{Helfrich},\g}=0,\quad c_0<0,
%\ee
%for any $\g\geq2$. Thus, (\ref{4.11}) is valid for all $\g=1,2,\dots$, collectively.

%In particular,  it is interesting to note that all the upper bounds above obtained for the Helfrich energy are independent of the value of $c_0$ but dependent on the sign of $c_0$ only.

The non-attainability of the infimum of the Helfrich energy (\ref{2}) among the ring tori suggests that it may be more realistic to study the minimization problem subject to
a fixed volume-to-area ratio, $\rho=\frac VA$, constraint, where $V$ may be proportional to the amount of energy needed for consumption of the living cell, and $A$
may serve to account for the amount of nutrients available to the cell through its contact with its environment. Recall that, for $\Sigma=T^2_{a,b}$, we have $V=2\pi^2 ab^2$ and $A=4\pi^2 ab$. So
we may set
\be
\rho=\frac b2=\mbox{ const}.
\ee
That is, the radius of the torus tube, $b$, is fixed, resulting in $a=\frac b\tau$. Substituting this result into (\ref{28}), we have
\be\label{4.13}
 U_{\mbox{Helfrich}}(T^2_{a,b})={2\pi^2\kappa}\left(\frac1{\sqrt{1-\tau^2}}-1+(bc_0 -1)^2\right)\frac1\tau\equiv 2\pi^2\kappa\,h_1(\tau),\quad\tau\in(0,1).
\ee
If $bc_0-1\neq0$, then $h_1(\tau)\to\infty$ as $\tau\to0$ and $\tau\to1$. Thus it may be shown that there is a unique $\tau_0\in(0,1)$ where $h_1(\tau)$ attains its minimum. In general, it is rather complicated to express $\tau_0$ in terms of the parameter $bc_0$. Here we only list a few simple cases: (i) The Willmore energy
case $c_0=0$. Then we have $\tau_0=\frac1{\sqrt{2}}$, which is exactly the Willmore ratio \cite{W,MN,W2,MN2}. (ii) The spontaneous curvature is three times 
of the constant principal curvature of the torus, $c_0=\frac3b$. We have
\be
\tau_0=\left(1-\left[\frac{(908+36\sqrt{633})^{\frac13}}{18}+\frac8{9(908+36\sqrt{633})^{\frac13}}-\frac29\right]^2\right)^{\frac12}\approx0.8491316067.
\ee
In general, let $c=(bc_0-1)^2$ in (\ref{4.13}). Then $\tau_0$ depends on $c>0$ monotonically such that $\tau_0\to0$ as $c\to 0$ and $\tau_0\to1$ as $c\to\infty$. If $bc_0-1=0$, then $c_0=\frac1b$. That is, the spontaneous curvature is equal to the constant principal curvature of the torus $T^2_{a,b}$. In this case the
function $h_1(\tau)$ coincides with $h(\tau)$ in (\ref{4.9}) so that the infimum of (\ref{4.13}) is zero which is not attainable.

In other words, subject to the fixed
volume-to-area ratio constraint, the Helfrich energy (\ref{2}) with $c_0\neq0$ has exactly one single isolated situation, when $c_0=\frac1b$, where the minimization of
the energy over embedded ring tori, $T^2_{a,b}$, does not have a solution. Otherwise, whenever $c_0\neq\frac1b$, the constrained minimization of the Helfrich energy (\ref{2})
over the same set of ring tori has a unique solution (with uniquely determined generating radii, $a$ and $b$). Thus, although the existence of an energy minimizer among the set of ring tori is
restored except for an isolated situation, the scale invariance breaks down.

We now consider the Helfrich bending energy in its general form \cite{H,ZH1} containing contributions from the volume and surface area of the lipid vesicle:
\be\label{416}
{\cal F}(\Sigma)=U_{\mbox{Helfrich}}(\Sigma)+p\int_{\cal V}\dd v+\lm\int_{\Sigma}\dd\sig,
\ee
where $\cal V$ is the solid region enclosed by the vesicle $\Sigma$, with $\dd v$ the volume element, and $p$ and $\lm$, respectively, are the osmotic pressure difference between the inside and outside of the lipid membrane and the surface tension. Physically, the lipid bilayer structure of the lipid membrane results in a one-way
traffic flow of salt, allowing salt to enter the cell but not leak away, which leads to a jump of salt concentration and hence a positive pressure difference, $p>0$.
On the other hand, surface tension of the plasma membrane of the cell dictates an elastic preference for the vesicle to assume as small a surface area as possible,
thus leading to $\lm>0$ as well. In our study below, we will observe these non-degenerate restrictions.

First, we let $\Sigma=S^2_R$ ($R>0$). We see that (\ref{416}) becomes ${\cal F}(S^2_R)=2\pi\kappa(c_0 R-2)^2+\frac{4\pi}3 R^3 p +4\pi R^2 \lm$. It is
clear that this function has no minimum when $c_0\leq0$ but it has a unique minimum when $c_0>0$ which is attained at
\be
R_0=\frac1{2p}\left(\sqrt{(2\lm+\kappa c_0^2)^2+8\kappa p c_0}-(2\lm+\kappa c_0^2)\right).
\ee
Next, we let $\Sigma=T^2_{a,b}$ ($a>b>0$). Using (\ref{28}), $V=2\pi^2 ab^2$, and $A=4\pi^2 ab$, we have ${\cal F}(T^2_{a,b})=2\pi^2 h_2(a,\tau)$, where
\be\label{4.18}
h_2(a,\tau)=\frac{\kappa}{\tau\sqrt{1-\tau^2}}-2\kappa c_0 a+\left(\kappa c_0^2+2\lm\right) a^2\tau+p a^3\tau^2.
\ee
It is clear that this function has no minimum when $c_0\leq0$. When $c_0>0$, we see that $\frac{\pa h_2}{\pa a}=0$ gives us the solution
\be\label{4.19}
a=\frac1{3p\tau}\left(\sqrt{(\kappa c_0^2+2\lm)^2+6\kappa c_0 p}-(\kappa c_0^2+2\lm)\right)\equiv\frac{b_0}\tau.
\ee
Inserting (\ref{4.19}) into (\ref{4.18}), we obtain
\be\label{4.20}
h_2\left(\frac{b_0}\tau,\tau\right)=\frac\kappa\tau\left(\frac{\tau^2}{\sqrt{1-\tau^2}(1+\sqrt{1-\tau^2})}+\beta_0\right)\equiv\kappa h_3(\tau),\quad \tau\in (0,1),
\ee
where
\be\label{412}
\beta_0=(b_0c_0-1)^2+\frac{b_0^2}\kappa(2\lm+p b_0)>0.
\ee
For any $\beta_0>0$, $h_3(\tau)$ has a unique minimizer for $\tau\in(0,1)$, which solves
$h_3'(\tau)=0$ or equivalently, the equation
\be\label{4.22}
\frac{1-2\tau^2}{(1-\tau^2)^{\frac32}}=1-\beta_0,\quad \tau\in(0,1).
\ee
We assert $\beta_0<1$. 
In fact, since $\frac{\pa h_2}{\pa a}=0$ gives us $3pb_0^2+2(\kappa c_0^2+2\lm)b_0-2\kappa c_0=0$, or
\be\label{414}
\frac{2p b_0^3}{\kappa}+\frac{b_0^2}{\kappa}(2\lm+pb_0)+2b_0^2 c_0^2-2b_0 c_0=0,
\ee
 we may insert (\ref{414}) into (\ref{412}) to get
\be\label{415}
\beta_0=1-\frac{2pb_0^3}{\kappa}-b_0^2 c_0^2<1,
\ee
as asserted.

Applying the conclusion that $\beta_0<1$ in (\ref{4.22}), we deduce that the unique solution $\tau_0$ of (\ref{4.22}) must satisfy the bound
\be\label{4.23}
0<\tau_0<\frac1{\sqrt{2}}.
\ee
With such $\tau_0$ and
(\ref{4.19}), we obtain the energy-minimizing ring torus $T^2_{a_0,b_0}$, which may be summarized as follows:
\be
a_0=\frac{b_0}{\tau_0},\quad b_0=\frac1{3p}\left(\sqrt{(\kappa c_0^2+2\lm)^2+6\kappa c_0 p}-(\kappa c_0^2+2\lm)\right).
\ee
It may be useful to note that  $\tau_0$  is a decreasing function of the right-hand side of (\ref{4.22}).

In summary, we see that the minimization of the energy (\ref{416}) over the set of spheres or the set of ring tori has a solution when
and only when the spontaneous curvature $c_0$ is positive.
Moreover,
in both cases, the radius and radii of the energy-minimizing sphere and torus, respectively, are uniquely and explicitly determined by the coupling parameters.
Furthermore, the ratio of the radii of the energy-minimizing torus universally lies in the open interval $\left(0,\frac1{\sqrt{2}}\right)$ in the entire parameter regime where
$c_0>0$. 

Thus, roughly speaking, the presence of the spontaneous curvature and surface-area and volume contributions to the bending energy, tunes down
the ratio of the generating radii of the energy-minimizing ring torus from that of the Willmore energy \cite{W,MN,W2,MN2} and breaks the scaling invariance of
the problem.

Note that (\ref{4.23}) may be used to obtain some refined estimates for $\tau_0$. For example, let $s=\frac1{\sqrt{1-\tau^2}}$ and use $B_0$ to denote the
right-hand side of (\ref{4.22}). Then $0<B_0<1$ and (\ref{4.22}) becomes $-s^3+2s=B_0$ with $1<s<\sqrt{2}$. Thus, from $s<s^2$ and $-s^2+2>0$, we have $s^2(-s^2+2)>s(-s^2+2)=B_0$, which leads to $s^2<1+\sqrt{1-B_0}=1+\sqrt{\beta_0}$. On the other hand, using $s^3-2s<s^2-2$, we get $s^2>2-B_0=1+\beta_0$. Consequently, returning to the variable $\tau$, we see that the ratio of the generating radii of the energy-minimizing torus, $\tau=\tau_0$, satisfies the estimates
\be\label{4.25}
\sqrt{1-\frac1{1+\beta_0}}<\tau_0<\sqrt{1-\frac1{1+\sqrt{\beta_0}}}.
\ee
It is of interest to notice that, the classical Willmore ratio, $\frac1{\sqrt{2}}$, would appear in the limiting situation $\beta_0\to1$ in (\ref{4.25}), but would
actually never happen
for any concrete choice of the coupling parameters.

It will be interesting to compare our results above on the minimization of the vesicle energy (\ref{416}) over the set of ring tori with those obtained in \cite{Zhong,Mutz}. In particular,  it is found therein that
the spontaneous curvature in the bending energy should stay positive for the existence of a stable toroidal vesicle, which is consistent with our results here.
(We note that in \cite{ZH1} the mean curvature is taken to be the negative of the average of the principal curvatures which effectively reverses the sign of
the spontaneous curvature in the Helfrich bending energy.)
However, their statement that their energy-minimizing torus has the same Willmore ratio, $\frac1{\sqrt{2}}$, for the generating radii, is inconsistent with our findings
shown above. These and other issues will be investigated and clarified in the last few paragraphs of the next section.

We may also consider the minimization of (\ref{416}) subject to a fixed  volume-to-surface-area ratio constraint over the set of embedded ring tori,
$\{T^2_{a,b}\}$. As seen earlier, the radius $b$ is now fixed, $b=b_0>0$, say. Thus, inserting $a=\frac{b_0}\tau$ into (\ref{4.18}), we  still obtain
(\ref{4.20}) and (\ref{412}), with $b_0$ and hence $\beta_0$ being arbitrary, though.
Therefore, now,   (\ref{4.22}) has a unique minimizer $\tau_0\in (0,1)$, which is given by 
\bea
\frac{2\tau_0^2-1}{(1-\tau_0^2)^{\frac32}}&=&\beta_0-1=b_0^2 c_0^2-2b_0 c_0+\frac{b_0^2}\kappa(2\lm+p b_0)\nn\\
&=&\left(b_0 c_0-1\right)^2-\left(1-\frac{b_0^2}\kappa(2\lm+pb_0)\right).
\eea
Thus, when $\frac{b_0^2}\kappa(2\lm+pb_0)>1$, we have $\tau_0>\frac1{\sqrt{2}}$ for all $c_0$; when  $\frac{b_0^2}\kappa(2\lm+pb_0)=1$, we have
$\tau_0=\frac1{\sqrt{2}}$ when $c_0=\frac1{b_0}$ and $\tau_0>\frac1{\sqrt{2}}$ otherwise; when  $\frac{b_0^2}\kappa(2\lm+pb_0)<1$, we have
\bea
\tau_0&=&\frac1{\sqrt{2}}\quad\mbox{when }
c_0=\frac1{b_0}\pm\frac1{b_0}\sqrt{1-\frac{b_0^2}\kappa(2\lm+pb_0)}\equiv\frac1{b_0}\pm C_0,\\
\tau_0&<&\frac1{\sqrt{2}}\quad\mbox{when }\frac1{b_0}-C_0<c_0<\frac1{b_0}+C_0,\\
\tau_0&>&\frac1{\sqrt{2}}\quad\mbox{when }c_0<\frac1{b_0}-C_0\mbox{ or }c_0>\frac1{b_0}+C_0.
\eea

Consequently, we see that, for any fixed volume-to-surface-area ratio and arbitrarily prescribed spontaneous curvature, the full Helfrich energy (\ref{416}) has a unique minimizer over the set
of embedded ring tori, with uniquely determined generating radii, such that, depending on the value of the spontaneous curvature, the ratio of the
generating radii may assume any value in the unit interval $(0,1)$, with the Willmore ratio $\frac1{\sqrt{2}}$ occurring at a few critical situations as
explicitly described above. In particular, we see that, now, the obstruction to the existence of an energy-minimizing torus, due to the presence of the spontaneous
curvature, completely disappears, although the scale invariance remains broken.

\section{The shape equation}
\setcounter{equation}{0}
\setcounter{figure}{0}

In this section, we derive the shape equation, which is the Euler--Lagrange equation associated with the anisotropic vesicle bending energy (\ref{3}) or (\ref{35}).

%For greater generality, we allow the total rigidity $\om$ and the rigidity discrepancy
%$\delta$ to be location-dependent quantities and rewrite (\ref{35}) as
%\be\label{35b}
%U(\sS)=\int_{\sS}\left(\om H^2-\frac12\om K\pm\delta H\sqrt{H^2-K}\right)\,\dd\sig.
%\ee

Use ${\bf x}: D\to \sS$, ${\bf x}={\bf x}(u,v)$, $(u,v)\in D\in\bfR^2$, to denote a local representation of the 2-surface $\Sigma$, and $E,F,G$, $e,f,g$ the matrix
entries of its first and second fundamental forms, respectively, such that
\bea\label{51}
E&=&{\bf x}_u\cdot{\bf x}_u, \quad F={\bf x}_u\cdot {\bf x}_v, \quad G={\bf x}_v\cdot{\bf x}_v,\nn\\
 e&=&{\bf x}_{uu}\cdot{\bf N}=-{\bf x}_u\cdot{\bf N}_u,\quad f={\bf x}_{uv}\cdot {\bf N}=-{\bf x}_u\cdot{\bf N}_v=-{\bf x}_v\cdot{\bf N}_u,\nn\\
 g&=&{\bf x}_{vv}\cdot{\bf N}=-{\bf x}_v\cdot{\bf N}_v,
\eea
where ${\bf N}$: $\Sigma\to S^2$ is the Gauss map. Consider the normal variation ${\bf x}^t(u,v)={\bf x}(u,v)+t w(u,v){\bf N}(u,v)$ where $w(u,v)$ is a scalar testing
function. For an ${\bf x}^t$-dependent quantity $Q$, we adopt the notation ${Q}^t=Q({\bf x}^t)$ and $\dot{Q}=\left(\frac{\dd }{\dd t}Q^t\right)_{t=0}$.
Thus, applying (\ref{51}), we have 
\be
\dot{F}=\dot{\bf x}_u\cdot{\bf x}_v+{\bf x}_u\cdot\dot{\bf x}_v=(w_u{\bf N}+w{\bf N}_u)\cdot{\bf x}_v + {\bf x}_u\cdot(w_v{\bf N}+w{\bf N}_v)=-2w f.
\ee
 Similarly,
$\dot{E}=-2 w e, \dot{G}=-2w g$. Recall also that the mean curvature $H$ and Gauss curvature $K$ are given by
\be\label{45}
H=\frac{eG-2fF+gE}{2(EG-F^2)},\quad K=\frac{eg-f^2}{EG-F^2}.
\ee
 Consequently, these lead to the well-known expression
\bea\label{dA}
\dot{|{\bf x}_u\times {\bf x}_v|}&=&\dot{\sqrt{EG-F^2}}\nn\\
&=&\frac1{2\sqrt{EG-F^2}}\left(\dot{E}G+E\dot{G}-2F\dot{F}\right)=-2H\sqrt{EG-F^2}\,w.
\eea
In addition, from $f^t=-{\bf x}^t_u\cdot{\bf N}_v^t$, we have 
\be\label{37}
\dot{f}=-\dot{\bf x}_u\cdot{\bf N}_v-{\bf x}_u\cdot \dot{\bf N}_v=-w{\bf N}_u\cdot{\bf N}_v-{\bf x}_u\cdot\dot{\bf N}_v.
\ee
Now, since $\dot{\bf N}$ is perpendicular to ${\bf N}$, we can express it as 
\be\label{38}
\dot{\bf N}=\alpha{\bf x}_u+\beta{\bf x}_v, 
\ee
which gives us
$
\alpha E+\beta F=\dot{\bf N}\cdot{\bf x}_u=-{\bf N}\cdot\dot{\bf x}_u=-w_u.
$
Similarly, we have $\alpha F+\beta G=\dot{\bf N}\cdot{\bf x}_v=-w_v$. Hence we obtain
\be
\alpha=\frac1{EG-F^2}\left(-G w_u+Fw_v\right),\quad \beta=\frac1{EG-F^2}\left(Fw_u-Ew_v\right).
\ee
Inserting (\ref{38}) into (\ref{37}), we obtain
\bea\label{40}
\dot{f}&=& -w{\bf N}_u\cdot{\bf N}_v-{\bf x}_u\cdot \left(\alpha_v{\bf x}_u+\beta_v{\bf x}_v+\alpha{\bf x}_{uv}+\beta{\bf x}_{vv}\right)\nn\\
&=&-w{\bf N}_u\cdot{\bf N}_v -\left(\alpha_v E+\beta_v F\right)-\alpha\left(\Gamma^u_{uv}E+\Gamma^v_{uv}F\right)-\beta\left(\Gamma_{vv}^u E+\Gamma^v_{vv}F\right),
\eea
where the $\Gamma$'s are the Christoffel symbols. Likewise, we also have
\bea
\dot{e}&=&-w{\bf N}_u\cdot{\bf N}_u-(\alpha_u E+\beta_u F)-\alpha(\Gamma^u_{uu}E+\Gamma^v_{uu}F)-\beta(\Gamma^u_{uv}E+\Gamma^v_{uv}F),\label{41}\\
\dot{g}&=&-w{\bf N}_v\cdot{\bf N}_v-(\alpha_v F+\beta_v G)-\alpha(\Gamma^u_{uv}F+\Gamma^v_{uv}G)-\beta(\Gamma^u_{vv}F+\Gamma^v_{vv}G).\label{42}
\eea

To calculate $\dot{e},\dot{f},\dot{g}$ with (\ref{40})--(\ref{42}), we need the useful identities
\bea
{\bf N}_u&=&\frac1{EG-F^2}\left([fF-eG]{\bf x}_u+[eF-fE]{\bf x}_v\right),\label{43}\\
 {\bf N}_v&=&\frac1{EG-F^2}\left([gF-fG]{\bf x}_u+[fF-gE]{\bf x}_v\right).\label{44}
\eea

Thus, by (\ref{45}), we have
\bea
\dot{H}&=&\frac1{2(EG-F^2)}\left(\dot{e}G+e\dot{G}-2\dot{f}F-2f\dot{F}+\dot{g}E+g\dot{E}\right)\nn\\
&&-\frac{H}{EG-F^2}\left(\dot{E}G+E\dot{G}-2F\dot{F}\right),\label{46}\\
\dot{K}&=&\frac1{EG-F^2}\left(\dot{e}g+e\dot{g}-2f\dot{f}\right)-\frac{K}{EG-F^2}\left(\dot{E}G+E\dot{G}-2F\dot{F}\right).\label{47}
\eea

In view of the results obtained for $\dot{E},\dot{F},\dot{G},\dot{e},\dot{f},\dot{g}$, we can use (\ref{46}) and (\ref{47}) to compute $\dot{H}$ and $\dot{K}$,
which are complicated.
In the following, we simplify our calculation by using the curvature coordinates, also known as lines of curvature \cite{Do,Struik}, so that $F=0, f=0$, which is valid to do near a non-umbilic point where
$k_1\neq k_2$ or $H^2-K>0$. In this situation, in view of (\ref{40})--(\ref{44}) and the correspondingly updated 
Christoffel symbols, 
\be
\Gamma_{uu}^u=\frac{E_u}{2E},\quad\Gamma_{uu}^v=-\frac{E_v}{2G},\quad \Gamma_{uv}^u=\frac{E_v}{2E},\quad \Gamma_{uv}^v=\frac{G_u}{2G},\quad \Gamma_{vv}^u=-\frac{G_u}{2E},\quad\Gamma_{vv}^v=\frac{G_v}{2G},
\ee
the coefficients $\alpha,\beta$, and the derivatives of the Gauss map $\bf N$, given by
\bea
\alpha&=&-\frac{w_u}E,\,\, \alpha_u=-\frac{w_{uu}E-w_u E_u}{E^2},\,\,\beta=-\frac{w_v}G,\,\,\beta_v=-\frac{w_{vv}G-w_v G_v}{G^2},\\
{\bf N}_u&=&-\frac eE{\bf x}_u,\,\, {\bf N}_v=-\frac gG{\bf x}_v,\,\,{\bf N}_u\cdot{\bf N}_u=\frac{e^2}E,\, \,{\bf N}_u\cdot{\bf N}_v=0,\,\,
{\bf N}_v\cdot{\bf N}_v=\frac{g^2}G,
\eea
we see that (\ref{41}) and (\ref{42}) become
\bea
\dot{e}&=&w_{uu}-\frac{w_u E_u}{2E}+\frac{w_vE_v}{2G}-w\frac{e^2}E,\label{516}\\
\dot{g}&=&w_{vv}-\frac{w_v G_v}{2G}+\frac{w_u G_u}{2E}-w\frac{g^2}G.\label{517}
\eea
Inserting (\ref{516}) and (\ref{517}) into (\ref{46}) and (\ref{47}), and using $k_1=\frac eE, k_2=\frac gG$, we arrive at the reduced expressions
\bea
\dot{H}&=&\frac1{2E}\left(w_{uu}-\frac{w_uE_u}{2E}+\frac{w_vE_v}{2G}\right)+
\frac1{2G}\left(w_{vv}-\frac{w_v G_v}{2G}+\frac{w_u G_u}{2E}\right)\nn\\
&&+(2H^2-K)w,\label{518}\\
\dot{K}&=&\frac{g}{EG}\left(w_{uu}-\frac{w_uE_u}{2E}+\frac{w_vE_v}{2G}\right)+\frac e{EG}\left(w_{vv}-\frac{w_v G_v}{2G}+\frac{w_u G_u}{2E}\right)
+2HK w.\quad\quad\label{519}
\eea
Now recall that the Laplace--Beltrami operator induced from the first fundamental form of the surface $\Sigma$ in terms of the chosen coordinates is given by
\be\label{520}
\Delta w=\frac1E w_{uu}+\frac1G w_{vv}+\frac1{2E}\left(\frac{G_u}{G}-\frac{E_u}{E}\right)w_u+\frac1{2G}\left(\frac{E_v}{E}-\frac{G_v}{G}\right) w_v.
\ee
Thus in view of (\ref{520}) we see that (\ref{518}) and (\ref{519}) take their suppressed form
\bea
\dot{H}&=&\frac12\Delta w+\left(2H^2-K\right)w,\label{49}\\
\dot{K}
&=&2H\Delta w-L w+2HK w,\label{50}
\eea
where 
\be
L w=\frac{k_1}E\left( w_{uu}-\frac{E_u}{2E} w_u+\frac{E_v}{2G}w_v\right)+
\frac{k_2}{G}\left(w_{vv}-\frac{G_v}{2G}w_v+\frac{G_u}{2E}w_u\right).
\ee
Furthermore, integrating by parts, we may shift away all derivatives on $w$ to get
 \be\label{53}
\int_\sS \eta Lw\,\dd\sig=\int_\sS \left(M\eta\right)w\,\dd\sig, 
\ee
where $M$ is the adjoint of the operator $L$ given by
\bea\label{54}
M\eta&=&\frac1{\sqrt{EG}}\left(\left[k_1\sqrt{\frac GE}\eta\right]_{uu}+\frac12\left[k_1\frac{E_u}E\sqrt{\frac GE}\eta\right]_u-\frac12\left[k_1\frac{E_v}{\sqrt{EG}}\eta\right]_v\right.\nn\\
&&\left.
\,\,\,\quad \quad+\left[k_2\sqrt{\frac EG}\eta\right]_{vv}+\frac12\left[k_2\frac{G_v}G\sqrt{\frac EG}\eta\right]_v-\frac12\left[k_2\frac{G_u}{\sqrt{EG}}\eta\right]_u\right).
\eea

Consequently, by (\ref{dA}), (\ref{49}), (\ref{50}), the self-adjointness of the operator $\Delta$, and (\ref{53}), we may obtain the Euler--Lagrange equation of the energy (\ref{3}) or (\ref{35}). Here, for greater generality and applicability, however, we consider instead the following shape energy,
extending (\ref{416}), as proposed in \cite{H}, along the line of the work
\cite{ZH1,ZH2}:
\bea\label{F}
{\cal F}(\sS)&=&U(\sS)+\xi_1\int_\sS k_1\,\dd\sig+\xi_2\int_\sS k_2\,\dd\sig+p\int_{\cal V}\,\dd v+\Lambda\int_\sS\,\dd\sig\nn\\
&=&U(\sS)+2{\xi}\int_\sS H\,\dd\sig\pm\zeta\int_\sS\sqrt{H^2-K}\,\dd\sig+p\int_{\cal V}\,\dd v+\Lambda\int_\sS\,\dd\sig,
\eea
where $\xi_1,\xi_2$ are quantities which may depend on the bending rigidities and spontaneous curvature of the lipid membrane,
${\xi}=\frac12(\xi_1+\xi_2)$, $\zeta=\xi_1-\xi_2$, with $k_{1,2}=H\pm\sqrt{H^2-K}$, and the Lagrange multipliers $p$ and $\Lambda$, respectively, 
collectively measure the osmotic pressure difference between the inside and outside of the lipid membrane and the surface tension, bending
rigidities, and spontaneous curvature
on the lipid membrane. Thus, in view of (\ref{dA}), (\ref{49}), and (\ref{50}), we arrive at the shape equation for the lipid membrane as the Euler--Lagrange equation
of the shape energy (\ref{F}):
\bea\label{55}
&& p-2\Lambda H+\om\Delta H \mp\frac12\,\Delta\left(\frac {\delta K+\zeta H}{\sqrt{H^2-K}}\right)\pm\frac12\, M\left(\frac {\delta H+\zeta}{\sqrt{H^2-K}}\right)-2H^2\left(\om H+2\xi \right)\nn\\
&&+2(\om H+\xi) (2H^2-K)
\pm\frac{\delta}{\sqrt{H^2-K}}\left(2H^4-3H^2 K+K^2\right)=0,
\eea
where the operator $M$ is defined as in (\ref{54}) in curvature coordinates. In the Helfrich isotropic limit \cite{H} where the leading term of the energy is given as in (\ref{2}), we have $\kappa_1=\kappa_2=\kappa, \delta=0,\xi_1=\xi_2=-\kappa c_0$ ($c_0$ being the spontaneous curvature), $\zeta=0$, $\om=2\kappa,\xi=-\kappa c_0$, and 
\be
\Lambda=\lm+\frac12\kappa c_0^2,
\ee
where $\lm$ is the surface tension of the lipid membrane. Thus the equation (\ref{55}) becomes the classical shape equation
\be\label{75}
p-2\lm H+\kappa(2H-c_0)(2H^2-2K+c_0 H)+2\kappa \Delta H=0,
\ee
as deduced in \cite{ZH1,ZH2}, whose limit with $p=0,\lm=0, c_0=0$ is the
Willmore equation \cite{MN,MN2}
\be\label{76}
\Delta H+2H(H^2-K)=0,
\ee
as should be anticipated. To compare (\ref{75}) and (\ref{76}), it may be instructive to rewrite (\ref{75}) in terms of the left-hand side of (\ref{76}) as
\be\label{5.33}
p-(2\lm+\kappa c_0^2)H+2\kappa c_0 K+2\kappa(\Delta H+2H[H^2-K])=0.
\ee
 Of course, these last three equations are coordinate-independent. Finally the Euler--Lagrange equation for the free energy (\ref{3}) may be obtained by setting $p=0,\Lambda=0, \xi_1=\xi_2=0$ in (\ref{55}),
which is, at a non-umbilical point where $H^2-K>0$,
\bea\label{77}
&&\om\Delta H \pm\frac\delta2\left( M\left[\frac {H}{\sqrt{H^2-K}}\right]-\Delta\left[\frac { K}{\sqrt{H^2-K}}\right]\right)
+2\om H(H^2-K)\nn\\
&&
\pm\frac{\delta}{\sqrt{H^2-K}}\left(2H^4-3H^2 K+K^2\right)=0,
\eea
which reduces to the Willmore equation (\ref{76}) again in the isotropic limit, $\delta=0$.

All these equations, (\ref{55}), (\ref{75}), (\ref{76}), and (\ref{77}) are fourth order.

We now study and clarify the issues raised in the preceding section with respect to our results there and those in \cite{Zhong}. For this purpose, we look for solutions
of the shape equation (\ref{75}) or (\ref{5.33}) among the set of ring tori, $\{T^2_{a,b}\}_{a>b>0}$. With the parametrization (\ref{3.7}), we have
$E={\bf x}_u\cdot{\bf x}_u=b^2, F={\bf x}_u\cdot{\bf x}_v=0, G={\bf x}_v\cdot{\bf x}_v=(a+b\cos u)^2$. Hence, in view of (\ref{11}), or
\be\label{5.35}
H=\frac1{2b}+\frac{\cos u}{2(a+b\cos u)},\quad K=\frac{\cos u}{b(a+b\cos u)},
\ee
and (\ref{520}), we get
\be\label{5.36}
\Delta H=-\frac{a}{2b^2(a+b\cos u)^3}\left(b+ {a\cos u}\right).
\ee
Substituting (\ref{5.35}) and (\ref{5.36}) into (\ref{5.33}), we arrive at the algebraic equation
\be\label{5.37}
\left(p-\frac{(2\lm+\kappa c_0^2)}b+\frac{2\kappa c_0}{b^2}\right)\xi^3+\frac ab\left(\lm+\frac{\kappa c_0^2}2-\frac{2\kappa c_0}b\right)\xi^2
-\frac{a\kappa}b\left(1-\frac{a^2}{2b^2}\right)=0,
\ee
where $\xi=a+b\cos u$. Setting all the coefficients of various powers of $\xi$ in the equation to zero, we obtain the {\em unique} solution
of (\ref{5.37}) given by the Willmore ratio $a=\sqrt{2}b$, and
\be\label{5.38}
p=\frac{2\kappa c_0}{b^2},\quad \lm=2\kappa c_0\left(\frac1b-\frac{c_0}4\right).
\ee
These formulas coincide with those derived in \cite{Zhong} when $b=1$. In particular, when $c_0=0$, we return to the Willmore equation, and the result shows that
the Willmore torus is the only solution of the Willmore equation in the set of ring tori. 

As an immediate consequence of our study in Section 4 and the above discussion, we conclude that, when $p,\lm>0$, the Helfrich shape equation (\ref{75}) or (\ref{5.33}) does not possess an energy-minimizing solution
in the set of ring tori for any value of the spontaneous curvature $c_0$.

The formulas in (\ref{5.38}) indicate that $p,\lm$ are {\em solution dependent}, but {\em not} simply determined by the other coupling parameters, $\kappa$ and $c_0$.
Furthermore, if we require $p,\lm>0$, then we need to impose the condition
\be\label{5.39}
0<c_0<\frac4b.
\ee
Of course, if no sign restriction is imposed to $p,\lm$, then the only solution of
the shape equation (\ref{5.33}) is the Willmore torus, $a=\sqrt{2}b$,  with $p,\lm$ satisfying (\ref{5.38}), for an arbitrarily given spontaneous curvature $c_0$.

We continue to assume  $p,\lm>0$ and aim to interpret (\ref{5.38}) differently. In fact,  practically, if $p,\lm$ are regarded as prescribed, then (\ref{5.38}) implies that $b$ is determined to be
\be\label{540}
b=\sqrt{\frac {2\kappa c_0}p}=\frac{4\kappa c_0}{2\lm+\kappa c_0^2},
\ee
which constrains the values of $p,\lm$ in view of $c_0$ and $\kappa$ by the relation
\be
p=\frac{(2\lm+\kappa c_0^2)^2}{8\kappa c_0}.
\ee
As a result, (\ref{5.39}) becomes
\be
c_0^{\frac32}<2\sqrt{\frac{2p}\kappa}.
\ee
All these are constraints imposed on the coupling parameters and $b$ appears as a parameter in the solution determined by these parameters. So there is no scale
invariance.

In the following, we evaluate the Helfrich bending energy (\ref{416}) for $\Sigma=T^2_{a,b}$ when $p,\lm$ are as given in (\ref{5.38}) (where $c_0$ is
arbitrary) or (\ref{540}) (where $c_0$ satisfies (\ref{5.39})). To do so, inserting
(\ref{5.38}) into (\ref{4.18}) and using $b=a\tau$ ($\tau\in(0,1)$), we see that (\ref{4.18}) becomes
\be\label{5.40}
h_2(a,\tau)=\frac{\kappa}{\tau\sqrt{1-\tau^2}}+4\kappa c_0 a=\frac{\kappa}{\tau\sqrt{1-\tau^2}}+\frac{4\kappa c_0 b}\tau.
\ee

Thus, if $a$ is fixed (but $b$ is changing along with $\tau$), then the minimum of $h_2(a,\tau)$ is attained at $\tau=\frac1{\sqrt{2}}$, as obtained in \cite{Zhong}.
However, here, there is no restriction to the sign of $c_0$.

On the other hand, if $b$ is fixed (so that $p,\lm$ are fixed), we see that the right-hand side of (\ref{5.40}) goes to $-\infty$ as $\tau\to0$ if
\be
c_0<-\frac1{4b}.
\ee
In particular, the energy has no global minimum over the set of ring tori. In the critical situation in which $c_0=\frac1{4b}$, it is seen that the right-hand side of
(\ref{5.40}) stays positive for $\tau\in(0,1)$ and tends to zero as $\tau\to0$. Hence, again, the energy has no global minimum over the set of ring tori. Finally, if
\be\label{5.42}
c_0>-\frac1{4b},
\ee
the right-hand side of (\ref{5.40}) blows up when $\tau\to0$. Therefore (\ref{5.40}) has a global minimum for $\tau\in(0,1)$ which may be found as the unique
root, say $\tau_0$, of the equation
\be\label{5.43}
2\tau^2-4c_0 b\,(1-\tau^2)^{\frac32}=1,\quad\tau\in(0,1),
\ee
since it can be examined that the uniqueness of the solution to (\ref{5.43}) is ensured by the condition $c_0>-\frac1{3b}$ which is contained in (\ref{5.42}).
It is clear that
\be\label{5.44}
\tau_0>\frac1{\sqrt{2}}\quad\mbox{if }c_0>0,\quad\tau_0<\frac1{\sqrt{2}}\quad\mbox{if }c_0<0,
\ee
and $\tau_0=\frac1{\sqrt{2}}$ only if $c_0=0$. Under the condition (\ref{5.42}), $\tau_0$ can be obtained explicitly. However, due to its complicated expression,
we omit presenting $\tau_0$ in terms of $c_0,b$ in its full generality but only list a few concrete simple results here:
\be\label{5.45}
\tau_0=\frac{\sqrt{3}}2,\quad c_0=\frac1b;\quad\tau_0=\frac{2\sqrt{2}}3,\quad c_0=\frac{21}{4b};\quad\tau_0=\frac{\sqrt{5}}3,\quad c_0=\frac3{32b},
\ee
all exceeding $\frac1{\sqrt{2}}$, and
\bea\label{5.46}
&&\tau_0=\sqrt{6\sqrt{5}-13},\quad c_0=-\frac{3}{32b};\quad\tau_0=\frac{\sqrt{7}}4,\quad c_0=-\frac2{27b};\nn\\
&&\tau_0=\frac{\sqrt{19-2\sqrt{29}}}7,\quad c_0=-\frac7{32b},
\eea
all staying below $\frac1{\sqrt{2}}$, as stated in (\ref{5.44}). Again, there is no restriction to the sign of $c_0$.

As another interpretation of (\ref{5.38}), we note that, for $\Sigma=T^2_{a,b}$, (\ref{5.38}) implies that the mean curvature $H$ and Gauss curvature $K$
satisfy the equation
\be
p-(2\lm+\kappa c_0^2)H+2\kappa c_0 \, K=0.
\ee
Consequently, under (\ref{5.38}), the shape equation (\ref{5.33}) implies the Willmore equation (\ref{76}). Hence the onset of the Willmore ratio,
$a={\sqrt{2}}b$, is hardly surprising.

Summarizing the above discussion related to the work \cite{Zhong}, we see that, when $a$ is fixed, the minimization of the Helfrich energy (\ref{416}) over
the set of ring tori $\{T^2_{a,b}\}$ with $p,\lm$ being given by (\ref{5.38}), is realized at the Willmore ratio, $\tau_0=\frac1{\sqrt{2}}$, without any sign restriction
for $c_0$; when $b$ is fixed, so that $p,\lm$ are fixed by (\ref{5.38}) as well, the problem of minimization of (\ref{416}) over $\{T^2_{a,b}\}$ has a solution if and
only if $c_0$ satisfies (\ref{5.42}), and the solution is given by a unique value $\tau_0$ of the ratio of the generating radii which may exceed or stay below $\frac1{\sqrt{2}}$ depending on whether $c_0$ is positive or negative, as stated in (\ref{5.44}). In both situations, the minimization is restricted, or partial, in
the sense that the parameters $p,\lm$ assume the form (\ref{5.38}), which is different from the unrestricted, global, minimization problem investigated in Section 4.
When $c_0\neq0$, the energy of the solution of the partial minimization problem  is below that of the Willmore solution, subject to (\ref{5.38}). In other words,
when $c_0\neq0$,
the Willmore solution, subject to (\ref{5.38}), cannot be a global energy minimizer, even over the set of ring tori.
\section{Conclusions and comments}
\setcounter{equation}{0}
\setcounter{figure}{0}

We have seen that  the presence of the spontaneous curvature $c_0$ in the Helfrich bending energy (\ref{2}) renders two difficulties: It makes the minimization of
the free energy
a no-go situation and it obstructs the derivation of  genus-dependent energy lower and upper bounds. In the former context, the difficulty is well demonstrated
for the minimization of the energy over the set of all embedded ring tori, $\{T^2_{a,b}\}_{a>b}$ (the minimization problem for genus-one surfaces). We have shown that the minimization of the free Helfrich energy (\ref{2})
over the set $\{T^2_{a,b}\}_{a>b}$ has no solution for any $c_0\neq0$. However, the same minimization problem has a unique solution subject to a fixed
volume-to-surface-area ratio constraint for the torus-shaped vesicles except for the single situation when $c_0$ is equal to the constant principal curvature of
the ring tori, $c_0=\frac1b$. In fact, even for the minimization problem for genus-zero surfaces, it fails to possess a solution among all spheres when $c_0<0$.
In the latter context, the difficulty arises as a consequence of the presence of a Gauss--Bonnet invariant, originating from
the integral of the Gauss curvature, which effectively conceals
the genus dependence of the energy.
In the present work we are able to overcome these two difficulties by using the scale-invariant curvature energy (\ref{3}) as the main free bending energy governing the shape of a lipid vesicle, and we arrive at the following conclusions.

\begin{enumerate}

\item[(i)] In sharp contrast to the situation of the Helfrich energy, the minimization problem of the curvature energy (\ref{3}) 
over the embedded ring tori $\{T^2_{a,b}\}_{a>b}$ always has a unique solution, up to rescaling of the generating radii, for arbitrary choice of the parameters, which recovers the Willmore solution,
$a=\sqrt{2}b$,
in the isotropic limit.

\item[(ii)] Anisotropy of the bending energy (\ref{3}) allows a broad range of phenomenology for the shapes of a vesicle and ellipsoidal and biconcave geometries 
may energetically be
favored over a spherical surface for a vesicle when the ratio of the difference and sum of bending rigidities lies in appropriate ranges. Numerical examples show that, as one makes anisotropy 
more and more significant, a transition process of surface shapes from spherical, to ellipsoidal, and then to biconcave geometries, is observed, under a fixed
volume-to-surface-area ratio constraint.

\item[(iii)] Genus-dependent topological lower and upper bounds are established for the bending energy (\ref{3}). Both bounds are linear in terms of the genus number of
the vesicle.

\item[(iv)] The bending energy (\ref{3}) enjoys a reinterpretation as the Helfrich free energy (\ref{2})  in which the spontaneous curvature is given as a location-dependent
quantity proportional to the difference of the principal curvatures of the vesicle.

\item[(v)] For the bending energy (\ref{3}), there are some anisotropic situations when toroidal surfaces are energetically favored over round spheres for a vesicle.

\item[(vi)] The Euler--Lagrange equation associated with the bending energy (\ref{3}) that governs the shape of a vesicle is derived in the general situation that recovers the classical
Helfrich shape equation in the isotropic limit.

\item[(vii)] Although the only solution of the Helfrich shape equation (\ref{75}) or (\ref{5.33}) over the set of ring tori is the Willmore torus
with $a=\sqrt{2}b$ and $b$ satisfying (\ref{540}), this solution is not a global energy minimizer over the set of ring tori.

\end{enumerate}

Of independent interest, we have also studied the problem of direct minimization of the Helfrich energy (\ref{416}), incorporating contributions from the surface area and volume of the 
vesicle. We conclude that, for the minimization over the set of spheres or the set of ring tori, an energy minimizer exists if and only if the spontaneous curvature is positive. Furthermore, the radius of the sphere and
radii of the torus that minimize the energy, over the respective sets, are all uniquely and explicitly determined by the coupling parameters in the bending energy. In addition, the ratio of
the radii of the energy-minimizing torus
among the ring tori stays in the universal interval $\left(0,\frac1{\sqrt{2}}\right)$, regardless of the values of the coupling parameters. It
is worth noting that such a parameter-independent estimate for the ratio of the ring-toroidal radii may be used for us to obtain some refined parameter-dependent estimates
for the ratio. Furthermore, we have carried out a study of the minimization of the Helfrich energy (\ref{416}) subject to a fixed volume-to-surface-area ratio constraint, over
the set of ring tori. We see that, now the spontaneous curvature obstruction to the existence of an energy-minimizing torus  completely disappears. In other words,
for any coupling parameters and spontaneous curvature, the minimization problem concerned has a unique least-energy torus solution. Moreover, the ratio of
the generating radii of the energy-minimizing torus may assume any value in the unit interval $(0,1)$, whether below, above, or at the Willmore ratio $\frac1{\sqrt{2}}$, depending on the ranges of the physical
and geometric parameters involved. In all these cases, there is a breakdown of the scale invariance.

This work also points to some problems of future interest. 

\begin{enumerate}

\item[(i)] We have seen in (\ref{6}) that the bending energy satisfies the absolute lower bound $U(\sS)\geq 4\pi\sqrt{\gamma}\kappa$ where $\gamma=\frac{\kappa_1}{\kappa_2}$ is the ratio of anisotropy and $\kappa=\kappa_2$. When $\gamma=1$, this lower bound is saturated by round spheres, as 
found by Willmore \cite{W,W2}. When $\gamma\neq1$, the lower bound should be realized by non-round spheres, whose geometric and topological properties would
depend on the value of $\gamma$. A determination of such a dependence may be intriguing but useful for vesicle phenomenology.

\item[(ii)] The upper bound stated in (\ref{20}), namely, $U_\g\leq 2\pi^2\g\kappa\frac{\tau(\gamma)}{(1-\tau^2(\gamma))^{\frac32}}$, $\g\geq1$, renders several questions of challenge 
to be answered. The simplest one may be when $\g=1$, which gives us the bound
\be\label{6.1}
U_1\leq 2\pi^2\kappa \frac{\tau(\gamma)}{(1-\tau^2(\gamma))^{\frac32}},\quad0<\gamma<\infty.
\ee
In the isotropic case when $\gamma=1$, we arrive at the Willmore problem, and it is shown \cite{MN,MN2}  that equality now holds in (\ref{6.1}). In
anisotropic situations in which $\gamma\neq1$,
however, we do not know whether equality holds in (\ref{6.1}), although along the lines of \cite{W,W2}  it may be reasonable to expect so as well.

\item[(iii)] The attainability of $U_\g$, $\g=0,1,2,\dots$, in (\ref{20}) is an open question except for the isotropic situation, $\gamma=1$, due to the work of Simon \cite{Si}. However, even in the isotropic situation in which $U_\g$ satisfies (\ref{21}), the value of $U_\g$ is unknown except for the two bottom cases, $\g=0,1$,
due respectively to Willmore \cite{W,W2} and Marques and Neves \cite{MN,MN2}
such that $U_0=4\pi\kappa$ and $U_1=4\pi^2\kappa$. For $\gamma=1$, some symmetry considerations suggest that the upper bound in (\ref{21}) 
when $\g\geq2$ might be further improved.

\item[(iv)] For the Helfrich energies (\ref{2}) and (\ref{416}) and the scale-invariant anisotropic curvature bending energy (\ref{3})
and the generalized energy (\ref{F}), it will be interesting to investigate the problem of energy minimization subject to
the constraints of  a fixed genus number and a fixed, say, volume-to-surface-area ratio of the vesicle.

\end{enumerate}

\medskip

{\bf Acknowledgments.} The author thanks Lei Cao for her careful reading and helpful comments on the first draft of this work.
This work was partially supported
by Natural Science Foundation of China under Grant No. 11471100.

\end{document}